\documentstyle[eqsecnum,preprint,prd,aps]{revtex}
\begin{document}\bibliographystyle{prsty}
\draft
\preprint{ADP-92-198/M13}
\title{
The Coherent State Representation of Quantum \\
Fluctuations in the Early Universe}
\author{A L Matacz\thanks{amatacz@physics.adelaide.edu.au }}
\address{
Department of Physics and Mathematical Physics, University of Adelaide\\
GPO Box 498, Adelaide, South Australia 5001}
\date{June 28, 1993}
\maketitle
\begin{abstract}
Using the squeezed state formalism the coherent state representation
of quantum fluctuations in an expanding universe is derived. It is shown
that this provides a useful alternative to the Wigner function as a
phase space representation of quantum fluctuations.
The quantum to classical
transition of fluctuations is naturally implemented by decohering the
density matrix in this representation. The entropy
of the decohered vacua is derived. It is shown
that the decoherence process breaks the physical equivalence between vacua
that differ by a coordinate dependent phase generated by a
surface term in the Lagrangian. In particular,
scale invariant power spectra are only obtained for a special choice of
surface term.
\end{abstract}
\pacs{98.80.Dr, 03.65.Fd}
\newpage
\section{Introduction}
One of the most attractive features of the Inflationary Universe scenario is
its ability to explain the origin of initial density perturbations
required to seed galaxies and galaxy clusters \cite{bra1}. During inflation,
initial quantum fluctuations of the ground state of the inflaton undergo
significant parametric amplification (squeezing) after Hubble crossing. This
leads to a macroscopic (i.e. many particle) quantum state.
In much of the previous work on this subject macroscopic was
incorrectly taken to be synonymous with classical thus the origin of classical
density perturbations was not properly addressed.
In actual fact the quantum state of the inflaton is spatially-homogeneous.
The assumption
typically made is that $<q_k^2>$ ($q_k$ is a
mode amplitude) can be interpreted as the amplitude of a
classical inhomogeneity. This can only be justified if the quantum state of
fluctuations is described by a statistical mixture of classical-like
spatially-inhomogeneous states. The transition of quantum fluctuations
from a pure spatially-homogeneous quantum state, to a statistical mixture of
spatially-inhomogeneous classical-like states,
can only occur via a decoherence process (from here on we shall refer to this
transition as simply the quantum to classical transition).

In order to get
decoherence it is necessary to go from a closed quantum system to an
open quantum system. One way to do this is via the introduction of an external
environment for the inflaton. By using simple toy model environments it has
been shown that decoherence in the coordinate representation is an effective
process on super-horizon scales \cite{dec}. However, as pointed out by
Laflamme and Matacz \cite{laf}, decoherence in the coordinate representation is
not always
a reliable criteria for the quantum to classical transition. Any realistic
model for an open system will introduce dissipation and fluctuation that
will greatly complicate and qualitatively change the dynamics of quantum
fluctuations \cite{hu2}. This will almost certainly have important
astrophysical implications for an inflationary phase. These implications
have not yet been addressed in the literature.

Given the complexity of a realistic open system,
it is worth looking at simple means of implementing the quantum to classical
transition. Recently
Brandenberger et al \cite{bra2} attempted to implement the quantum to
classical transition by
decohering the quantum state of fluctuations in the number state
representation. Gasperini and Giovannini \cite{gas} have implemented a scheme
which decoheres in the basis of what they call the superfluctuant operator.
These authors were interested in calculating the the entropy of
cosmological perturbations.
They utilized the squeezed state formalism and, with these
coarse graining schemes, obtained
the same entropy in the high squeezing limit.

The adoption of the language of squeezed states to cosmological particle
creation was first introduced by Grishchuk and Sidorov \cite{gri}.
Albrecht et al \cite{alb} have pointed
out that the squeezed state formalism contains no new physics in itself.
In fact, as noted by Hu et al \cite{hu1}, (who have used the squeezed state
formalism to discuss the role of intitial states in particle creation and
fluctuation in particle number) the
squeeze and rotation operators were derived, based on earlier work by
Kamefuchi and Umezawa \cite{kam}, in
Parker's original work on cosmological particle creation \cite{par}. Although
the physics is not new, the squeezed state
formalism gives an alternative description which can draw upon developments
in quantum optics. It is valid for any system described by a time dependent
quadratic Hamiltonian. Thus it could describe scalar fields, gravitons or
gauge invariant cosmological perturbations.

In this paper I have made use of the squeezed state formalism to derive
the coherent state representation (CSR) of quantum fluctuations in an
expanding FRW universe. This idea stems from work in
quantum optics which has shown that many states, including squeezed
states, can be represented as one dimensional superpositions over coherent
states \cite{fol}.
As is well known coherent states \cite{lou} describe classical-like,
spatially-inhomogeneous quantum states since they have well defined amplitude
and momentum. Thus they are the best quantum analogue of points in phase space.
The Wigner function has previously been used as a phase space representation
of quantum fluctuations in an expanding FRW universe \cite{hal}\cite{gri}.
In general
the Wigner function shows oscillatory behaviour and associated
negative regions. For these reasons it can not be considered a true phase space
probability distribution. It is accepted that these properties
are the signature of non-classical quantum interference effects \cite{zur}.
However, the Wigner function of a gaussian state, like the squeezed vacuum, is
a positive definite gaussian. This may lead one to incorrectly suspect that
squeezed vacua can be thought of as classical-like states.
The advantages
of the CSR over the Wigner function is that it shows explicitly how squeezed
quantum fluctuations are built from quantum superpositions over coherent
states. This is of great pedagogical value
in understanding the difference between quantum and classical fluctuations and
hence the need for decoherence.
Like the Wigner function, phase space information is also included since each
coherent state with
support in the superposition has a well defined amplitude and momentum.
More importantly, decohering the squeezed vacuum in the CSR provides a simple
and, as discussed below, a physically well motivated
means of implementing the quantum to classical transition of fluctuations.

Studies of environmentally induced decoherence \cite{zur2} have shown that
coherent states
are the most robust to the effects of a dissipative environment.
This singles out the coherent state basis as a preferred basis for decoherence.
For the case of scattering or non-dissipative environments, we would
expect decoherence to be most effective in a number state basis \cite{wal}.
However, in the early
universe we expect environments to be dissipative \cite{hu2}.
Decoherence in the CSR is therefore a well justified alternative to the
decoherence schemes advocated in \cite{bra2}\cite{gas}.
Decoherence in a CSR is a desirable result since it implies the
transformation of a coherent quantum phase space distribution to an
incoherent classical phase space distribution. Such a process is necessary
before we can, as
Grishchuk and Sidorov advocated \cite{gri}, adopt and interpret a squeezed
vacuum as a classical stochastic collection of standing waves.

The expectation
values of observables calculated using decohered vacua will in general be
different to that calculated using the corresponding pure states.
I will also
show that decoherence
breaks the physical equivalence between vacua that differ by
a coordinate dependent phase generated by surface terms in the
Lagrangian. It is
obviously important to see if the loss of quantum coherence greatly changes
the basic predictions of the pure states. In this paper I calculate the
entropy and the
amplitude and momentum fluctuations for two vacua defined with and without
a surface term in the Lagrangian. I explicitly solve and compare results
for the mixed and pure states of both vacua, in the after-Hubble crossing,
or high-squeezing limit of a de Sitter phase. Finally, I discuss the important
implications of these results to the power specta of fluctuations generated
from inflation.

\section{the model}
In this section I will show how a general real scalar field in an expanding
FRW universe is reduced to a quadratic time dependent Hamiltonian.
It also applies to the case of gravity wave perturbations which are equivalent
to the massless, minimally coupled scalar field (see \cite{grish} for
details). In section 6 we show how our results for the scalar field can be
applied to gauge invariant cosmological perturbations.

The action for a free scalar field in an arbitrary space-time can be written as
\begin{equation}
S=\int {\cal L}(x)d^4x= \int\frac{\sqrt{-g}}{2}d^4x
\left(g^{\mu\nu}\bigtriangledown_{\mu}\Phi\bigtriangledown_{\nu}\Phi-
(m^2+\xi  R)\Phi^2\right).
\end{equation}
In the spatially flat expanding metric
\begin{equation}
ds^2=a^2(\eta)[d\eta^2-\sum dx^2_i]
\end{equation}
we can write
\begin{equation}
{\cal
L}(x)=\frac{1}{2}a^2(\eta)\left[(\dot{\Phi})^2-\sum_{i}(\Phi_{,i})^2-\left(m^2
a^2+6\xi  \frac{\ddot{a}}{a}\right)\Phi^2\right]
\end{equation}
where a dot denotes a derivative with respect to conformal time.
If we rescale the field variable $\chi=a\Phi$, (2.3) then becomes
\begin{equation}
{\cal L}(x)=\frac{1}{2}\left[(\dot{\chi})^2-\sum_{i}(\chi_{,i})^2-
\left(m^2 a^2+(6\xi -1)\frac{\ddot{a}}{a}\right)\chi^2-
(1-6\xi )\frac{d}{d\eta}\left(\frac{\dot{a}}{a}\chi^2\right)\right]
\end{equation}
where the final term in (2.4) is the surface term. The part of the surface term
proportional to $\xi $ has been added in by hand. The surface term in (2.4)
ensures that the second derivative of the scale factor doesn't appear in the
Lagrangian. This is necessary to have a consistent variational theory when
the scale factor is treated dynamically rather than kinematically \cite{sim}.
However, despite this, the surface term is often dropped when the scale factor
is kinematic.
In this paper we will consider two cases: with and without the surface term.
All quantities
derived where the surface term has been kept will be denoted with an $s$
subscript.

We can expand the scalar field in a box of co-moving volume $L^3$ (fixed
coordinate volume)
\begin{equation}
\chi(x)=\sqrt{\frac{2}{L^3}}\sum_{\vec{k}}[q_{\vec{k}}^+
 \cos\vec{k}\cdot\vec{x} + q_{\vec{k}}^- \sin\vec{k}\cdot\vec{x}]
\end{equation}
which leads to the Lagrangians
\begin{eqnarray}
L(\eta)&=&\frac{1}{2}\sum_{\sigma}^{+-}\sum_{\vec{k}}\left[(\dot{q}_{\vec{k}}
^{\sigma})^2-\left(k^2+m^2 a^2+(6\xi -1)\frac{\ddot{a}}{a}
\right)q_{\vec{k}}^{\sigma2}\right] \\
L_s (\eta)&=&\frac{1}{2}\sum_{\sigma}^{+-}\sum_{\vec{k}}
\left[(\dot{q}_{\vec{k}}^{\sigma})^2-2(1-6\xi )\frac{\dot{a}}{a}
q_{\vec{k}}^{\sigma}\dot{q}_{\vec{k}}^{\sigma}
-\left(k^2+m^2 a^2+(6\xi -1)\frac{\dot{a}^2}{a^2}
\right)q_{\vec{k}}^{\sigma2}\right]
\end{eqnarray}
where k=$|\vec{k}|$ and $L(\eta)=\int{\cal L}(x)d^3 \vec{x}$.
Canonical momenta are
\begin{eqnarray}
p_{\vec{k}}^{\sigma}&=&\frac{\partial L(\eta)}{\partial\dot{q}_{\vec{k}}^
{\sigma}}=\dot{q}_{\vec{k}}^{\sigma} \\
p_{s\vec{k}}^{\sigma}&=&\frac{\partial L_s (\eta)}{\partial\dot{q}_{\vec{k}}^
{\sigma}}=\dot{q}_{\vec{k}}^{\sigma}-(1-6\xi
)\frac{\dot{a}}{a}q_{\vec{k}}^{\sigma}.
\end{eqnarray}
Defining the canonical Hamiltonian the usual way we find
\begin{equation}
H(\eta)=\frac{1}{2}\sum_{\sigma}^{+-}\sum_{\vec{k}}\left[p_{\vec{k}}^{\sigma
2}+\left(k^2+m^2 a^2+(6\xi -1)\frac{\ddot{a}}{a}\right)
q_{\vec{k}}^{\sigma2}\right]
\end{equation}
\begin{eqnarray}
H_s (\eta)&=&\frac{1}{2}\sum_{\sigma}^{+-}\sum_{\vec{k}}
\left[p_{s\vec{k}}^{\sigma 2}+(1-6\xi )\frac{\dot{a}}{a}(p_{s\vec{k}}
^{\sigma}q_{\vec{k}}^{\sigma}+q_{\vec{k}}^{\sigma}p_{s\vec{k}}^{\sigma})\right.
\nonumber \\ &+&
\left.\left(k^2+m^2 a^2+6\xi (6\xi -1)\frac{\dot{a}^2}{a^2}\right)q_{\vec{k}}
^{\sigma 2}\right]
\end{eqnarray}
where the sum is over positive $k$ only since we have an expansion
over standing rather than travelling waves.

The system is quantized by promoting $(p_{\vec{k}}^{\sigma},q_{\vec{k}}
^{\sigma}),\;(p_{s\vec{k}}^{\sigma},q_{\vec{k}}
^{\sigma})$
to operators obeying the usual harmonic oscillator commutation relation.
Thus the dynamics is reduced to the dynamics of time-dependent harmonic
oscillators. The Hamiltonian is not unique and is a result of our choice of
canonical variables. As equations (2.8-9) show, dropping the surface term is
the same as a
canonical transformation that only changes the canonical momentum.
We see from equations (2.10-11) that this leads to two different
Hamiltonians which inturn will define two vacua which
are different up to a coordinate dependent phase. We can show this as follows.
Consider the following addition of a general surface term to a Lagrangian
\begin{eqnarray}
\bar{L}(q,\dot{q})&\rightarrow& L(q,\dot{q})-
\frac{d}{dt}f(q,t) \nonumber \\
&\rightarrow& L(q,\dot{q})-\frac{\partial f}{\partial q}\dot{q}-\frac{\partial
f}{\partial t}.
\end{eqnarray}
This is the same as a point transformation on the Lagrangian. This
transformation changes the canonical momentum to
\begin{equation}
\bar{p}=\frac{\partial \bar{L}}{\partial \dot{q}}\rightarrow
p-\frac{\partial f}{\partial q}
\end{equation}
where $p$ is the canonical momentum of the original Lagrangian.
{}From (2.12) we find that the action transforms as
\begin{equation}
\bar{S}[q]\rightarrow S[q] -f(q_f,t_f)+f(q_i,t_i).
\end{equation}
This point transformation doesn't affect the classical equation of motion
because they are derived from the stationary action condition
$\delta S=S[q(t)]-S[q(t)+\delta q(t)]=0$ where $\delta q(t)$ vanishes at the
endpoints.
However from the general expression $U(q_f,t_f;q_i,t_i)=
N\sum_{paths}e^{iS}$ for the quantum propagator we can see that under
the transformation (2.14) the quantum propagator transforms as
\begin{equation}
\bar{U}(q_f,t_f;q_i,t_i)\rightarrow e^{-if(q_f,t_f)}
U(q_f,t_f;q_i,t_i) e^{if(q_i,t_i)}
\end{equation}
which inturn means that the wavefunction transforms as
\begin{equation}
\bar{\psi}(q,t)\rightarrow e^{-if(q,t)}\psi(q,t).
\end{equation}

The effect of this phase on average values is as follows. Consider the
observable $g(q,p)$. The average value of this observable with respect
to the transformed wavefunction (2.16) is
\begin{equation}
<g(q,p)>=\int dq e^{if(q,t)}\psi^*(q,t)g(q,p)e^{-if(q,t)}\psi(q,t).
\end{equation}
Obviously if $g$ is only a function of $q$ then everything commutes
and the phase cancells. When $g$ is also a function of $p$ we must write,
using (2.13)
\begin{equation}
p=-i\frac{\partial}{\partial q}+\frac{\partial f}{\partial q}
\end{equation}
remembering that now $\bar{p}=-i\frac{\partial}{\partial q}$ since it is the
new canonical momentum. We therefore have
\begin{equation}
pe^{-if(q,t)}\psi(q,t)=-ie^{-if(q,t)}\frac{\partial \psi(q,t)}{\partial q}.
\end{equation}
Clearly then the phase in (2.17) will cancell in general
and it therefore has
no effect on the expectation values of observables.
Thus vacua which differ by a coordinate dependent phase are considered
physically equivalent.
However, as will be shown in this paper, decoherence breaks this equivalence.

Changing
the time coordinate or rescaling the field variables are also a form of time
dependent canonical transformation. These canonical transformations also
change the form of the Hamiltonian which corresponds to selecting
different vacuum states.
We have followed the results of \cite{wei} which suggest
that using the rescaled field and conformal time is a preferred procedure.

\section{propagator for a generalised harmonic oscillator}

Consider the generalised harmonic oscillator defined by the Hamiltonian
\begin{equation}
\hat{H}(\eta)=b_1(\eta)\frac{\hat{p}^2}{2}+b_2(\eta)
\frac{(\hat{p}\hat{q}+\hat{q}\hat{p})}{2}+b_3(\eta)\frac{k^2 \hat{q}^2}{2}
\end{equation}
where $[\hat{q},\hat{p}]=i$. We define creation and annihilation
operators as
\begin{equation}
\hat{a}=\frac{\gamma \hat{q}+i\hat{p}}{\sqrt{2\gamma}},\;\;\;\hat{a}^{\dag}
=\frac{\gamma \hat{q}-i\hat{p}}{\sqrt{2\gamma}}
\end{equation}
where $\gamma$ is an arbitrary positive real.
The Hamiltonian can be written in the form
\begin{equation}
\hat{H}(\eta)=f(\eta)\hat{A}+f^*(\eta)\hat{A}^{\dag}+h(\eta)\hat{B}
\end{equation}
where
\begin{equation}
f(\eta)=\frac{b_3 k^2}{2\gamma}-ib_2-\frac{b_1 \gamma}{2},\;\;\;\;\;
h(\eta)=\frac{b_3 k^2}{2\gamma}+\frac{b_1 \gamma}{2}
\end{equation}
and
\begin{equation}
\hat{A}=\frac{\hat{a}^2}{2},\;\;\;\;\;\hat{A}^{\dag}=
\frac{\hat{a}^{\dag 2}}{2},\;\;\;\;\;\hat{B}=\hat{a}^{\dag}\hat{a}+1/2.
\end{equation}

We want to find the propagator for (3.1). To do this we make the the ansatz
\begin{equation}
\hat{U}(\eta,\eta')=e^{x(\eta)\hat{B}}e^{y(\eta)\hat{A}}e^{z(\eta)\hat{A}
^{\dag}}.
\end{equation}
It must satisfy the evolution equation for the propagator
\begin{equation}
\hat{H}(\eta)\hat{U}(\eta,\eta')=i\frac{\partial}{\partial
\eta}\hat{U}(\eta,\eta')
\end{equation}
subject to the initial condition $\hat{U}(\eta',\eta')=1$.
We find that the operators $\hat{A}, \hat{A}^{\dag}, \hat{B}$
satisfy the closed algebra
\begin{equation}
[\hat{A},\hat{A}^{\dag}]=\hat{B}=\hat{B}^{\dag},\;\;\;\;\;[\hat{A},\hat{B}]=
2\hat{A},\;\;\;\;\;[\hat{A}^{\dag},\hat{B}]=-2\hat{A}^{\dag}.
\end{equation}
Making use of the above closed algebra and the operator relation
\begin{equation}
e^{u\hat{O}}\hat{P}e^{-u\hat{O}}=\hat{P}+u[\hat{O},\hat{P}]+\frac{u^2}{2!}[\hat{O},[\hat{O},\hat{P}]]+...
\end{equation}
we find that
\begin{equation}
e^{x\hat{B}}\hat{A}=e^{-2x}\hat{A}e^{x\hat{B}},\;\;\;\;\;
e^{y\hat{A}}\hat{A}^{\dag}=(\hat{A}^{\dag}+\hat{B}y+y^2
\hat{A})e^{y\hat{A}},\;\;\;\;\;e^{x\hat{B}}\hat{A}^{\dag}=
e^{2x}\hat{A}^{\dag}e^{x\hat{B}}.
\end{equation}
Substituting (3.6) into (3.7) and using (3.3 and 3.10) we obtain the following
system of equations
\begin{eqnarray}
-if&=&\dot{y}e^{-2x}+\dot{z}y^2e^{-2x} \\
-if^*&=&\dot{z}e^{2x} \\
-ih&=&\dot{x}+\dot{z}y
\end{eqnarray}
which we must solve subject to the initial conditions $x(\eta'), y(\eta'),
z(\eta')=1$.

As it stands (3.6) is not necessarily unitary. Thus $x,y,z$ must satisfy some
further restrictions. If we write
\begin{equation}
x=ln\alpha,\;\;\;\;\;y=-\beta\alpha,\;\;\;\;\;z=\beta^*/\alpha
\end{equation}
where
\begin{equation}
\alpha=e^{-i\theta}coshr,\;\;\;\;\;\beta=-e^{-2i\varphi}sinhr
\end{equation}
then we can write (3.6) as
\begin{equation}
\hat{U}(\eta,\eta')=\hat{S}(r,\phi)\hat{R}(\theta)
\end{equation}
where $\phi=\varphi-\theta/2$ and
\begin{equation}
\hat{R}(\theta)=e^{-i\theta \hat{B}},\;\;\;\;\;
\hat{S}(r,\phi)=\exp [r(\hat{A}e^{-2i\phi}-\hat{A}^{\dag}e^{2i\phi})].
\end{equation}
$\hat{S}$ and $\hat{R}$ are called squeeze and rotation operators
respectively \cite{sch}. They are both unitary as is required.

The interesting property of a squeeze operator is
that it squeezes fluctuations in one quadrature at the expense of the other.
{}From the properties
\begin{equation}
\hat{S}^{\dag}\hat{a}^{\dag}\hat{S}=\hat{a}^{\dag}\cosh r
-\hat{a}e^{-2i\phi}\sinh r
\end{equation}
\begin{equation}
\hat{S}^{\dag}\hat{a}\hat{S}=\hat{a}\cosh r- \hat{a}^{\dag}e^{2i\phi}\sinh r
\end{equation}
we can derive the fundamental properties of a squeezed vacuum state
$\hat{S}(r,\phi)|0\rangle$ which are
\begin{eqnarray}
<\hat{q}^2>&=&\frac{1}{2\gamma}[\cosh^2 r+\sinh^2 r-2\cos 2\phi\cosh r\sinh r]
\\
<\hat{p}^2>&=&\frac{\gamma}{2}[\cosh^2 r+\sinh^2 r+2\cos 2\phi\cosh r\sinh r]
\\
<\hat{q}\hat{p}&+&\hat{p}\hat{q}>=
-2\sin 2\phi\cosh r\sinh r.
\end{eqnarray}
The squeeze parameter $r$ determines the strength of the squeezing
while the squeeze angle $\phi$ determines the distribution of the squeezing
between conjugate variables. We note that the lower bound of the uncertainty
relation is satisfied only when $\phi=n\pi/2$.

Substituting (3.14) into (3.11-3.13) we arrive at the neater
equations
\begin{eqnarray}
\dot{\alpha}&=&-if^*\beta-ih\alpha \\
\dot{\beta}&=&ih\beta+if\alpha.
\end{eqnarray}
If we are only interested in the vacuum state rather than the
complete propagator it may be better to reduce the above system to a single
second order differential equation.
We can do this as follows. Putting
\begin{equation}
\mu=\frac{\beta^*-\alpha^*}{\beta^*+\alpha^*}
\end{equation}
we find that, using (3.23-24)
\begin{equation}
2\dot{\mu}-i\mu^2(2h-f-f^*)+2i\mu(f-f^*)+i(f+f^*+2h)=0.
\end{equation}
Substituting
\begin{equation}
\mu=\frac{2i}{2h-f-f^*}\frac{\dot{g}}{g}
\end{equation}
we find
\begin{equation}
\ddot{g}+\dot{g}\left(\frac{\dot{f}+\dot{f^*}-2\dot{h}}{2h-f-f^*}+
i(f-f^*)\right)+\frac{4h^2-(f+f^*)^2}{4}g=0.
\end{equation}
We can rewrite (3.27-28) as
\begin{equation}
\mu=\frac{i}{\gamma b_1}\frac{\dot{g}}{g}
\end{equation}
and
\begin{equation}
\ddot{g}+\dot{g}(2b_2-\dot{b_1}/b_1)+k^2b_1b_3g=0.
\end{equation}

We require that $r(\eta')=0$ so we must choose our solution of (3.30) so
that $\mu(\eta')=-1$. In most cases we will choose $\gamma$ so that $f(\eta')$
in (3.4) will vanish.
{}From (3.15) and (3.25)
\begin{equation}
\mu=\frac{1+e^{2i\phi}\tanh r}{-1+e^{2i\phi}\tanh r}=
\frac{-1+\tanh^2 r -2i\sin 2\phi \tanh r}{1+\tanh^2 r -2\cos 2\phi\tanh r}.
\end{equation}
The solution of (3.30) therefore determines the squeeze operator.

Using (3.31) we can determine the squeeze parameter  from
\begin{equation}
\tanh^2 r=\frac{1+\mu+\mu^*+|\mu|^2}{1-\mu-\mu^* +|\mu|^2}.
\end{equation}
Given the squeeze parameter we can then, using (3.31), solve for $\sin 2\phi$
and
$\cos 2\phi$.
To solve for the rotation operator we use (3.25) and (3.23) and find that
\begin{equation}
\frac{\dot{\alpha}}{\alpha}=-if^*\frac{1+\mu^*}{1-\mu^*}-ih.
\end{equation}
This is solved by
\begin{equation}
\alpha(\eta)=\exp\left[-i\int_{\eta'}^{\eta}dt(f^*(1+\mu^*)/(1-\mu^*)+h)\right].
\end{equation}
Using (3.15) we can then write
\begin{equation}
\theta(\eta)=\frac{1}{2}\int_{\eta'}^{\eta}d\eta\left(2h+f^*\frac{(1+\mu^*)}
{1-\mu^*}+f\frac{(1+\mu)}{1-\mu}\right).
\end{equation}
A similar procedure with (3.24) gives
\begin{equation}
2\varphi=-\frac{1}{2}\int^{\eta}d\eta\left(2h+f^*\frac{(1-\mu^*)}
{1+\mu^*}+f\frac{(1-\mu)}{1+\mu}\right)+2\varphi_c.
\end{equation}
The constant contribution to the phase $\theta(\eta)$ is determined by the
requirement that $\theta(\eta')=0$. We do not require $2\varphi(\eta')=0$.
Thus $2\varphi_c$ must be chosen carefully so that the equations of motion
(3.23-24)
are satisfied.
For the rest of the paper we shall deal only with the squeezed vacuum
$|r,\phi\rangle=e^{-i\theta/2}S(r,\phi)|0\rangle$, though clearly we have a
formalism which can deal with more general initial states.

The squeezed vacuum has the coordinate space representation \cite{hon}
\begin{equation}
\psi_{r,\phi}(q)=e^{-i\theta/2}\left(\frac{\gamma}{\pi}\right)^{1/4}
(\cosh r-e^{2i\phi}\sinh r)^{-1/2}\exp\left[\frac{-\gamma q^2}{2}
\left(\frac{1+e^{2i\phi}\tanh r}{1-e^{2i\phi}\tanh r}\right)\right].
\end{equation}
The term in the curved brackets in the exponential is nothing but $-\mu$
defined in (3.29).
This is the usual way of studying quantum fluctuations in the Schrodinger
picture \cite{bra4}\cite{rat}. Thus the wavefunction (3.37) and (3.29-30) show
the necessary equivalence between the squeezed state formalism and the
coordinate representation methods.

Albrecht
et al \cite{alb} have also derived the equations of motion for the squeeze
paramater $r$,
squeeze angle $\phi$ and the phase $\theta$. Their equations are three coupled
first order nonlinear equations. On the other hand the equations  derived
here (3.23-24) are two coupled first order linear equations. These equations
were previously derived by Fernandez \cite{fer} using a different procedure.
The interested  reader is referred there for
other references dealing with time dependent quadratic Hamiltonians.

\section{The Coherent State Representation}
As is well known, coherent states \cite{lou} describe classical-like
states since they have well defined amplitude and momentum. Therefore they are
the best quantum analogue of points in phase space.
For these reasons the CSR is well
suited to highlighting the difference between quantum and classical
fluctuations.

Recent work
motivated by quantum optics has shown how squeezed
states can be represented as one dimensional superpositions over coherent
states \cite{fol}.
In this representation the squeezed vacuum has the form
\begin{equation}
|r,\phi\rangle=e^{-i\theta/2}(2\pi\sinh r)^{-1/2}\int_{-\infty}^{\infty}
\exp \left[-y^2\left(\frac{1-\tanh r}{2\tanh r}\right)\right]
|-iy e^{i\phi}\rangle dy
\end{equation}
where the expansion is over coherent states defined as
eigenstates of the annihilation operator,
$\hat{a}|\alpha\rangle=\alpha|\alpha\rangle$ where $\alpha$ is complex.
These Gaussian states are minimum uncertainty packets in $\hat{q}$ and
$\hat{p}$ with mean values determined by
\begin{equation}
\alpha=\frac{1}{\sqrt{2\gamma}}(\gamma<q>+i<p>).
\end{equation}
The mean values for the coherent states with
support in the superpositions are therefore determined by
\begin{equation}
-iy e^{i\phi}=\frac{1}{\sqrt{2\gamma}}(\gamma<q>+i<p>).
\end{equation}

When written as a density matrix (4.1) becomes
\begin{equation}
\hat{\rho}=\frac{1}{2\pi \sinh r}\int_{-\infty}^{\infty}
\exp \left[-(y^2+y'^2)\left(\frac{1-\tanh r}{2\tanh r}\right)\right]|-iy
e^{i\phi}\rangle\langle -iy'e^{i\phi}|dydy'.
\end{equation}
Dropping the off-diagonal terms in this representation corresponds
to decohering the squeezed vacuum in phase space. The resulting normalised
density matrix is
\begin{equation}
\hat{\rho}=\left(\frac{1-\tanh r}{\pi\tanh r}\right)^{1/2}
\int_{-\infty}^{\infty}
\exp \left[-y^2\left(\frac{1-\tanh r}{\tanh r}\right)\right]|-iy
e^{i\phi}\rangle\langle -iye^{i\phi}|dy.
\end{equation}

We find that mean values with respect to the decohered squeezed vacuum (4.5)
are
\begin{equation}
<\hat{q}^2>_m=\frac{1-\cos 2\phi\tanh r}{2\gamma(1-\tanh r)}
\end{equation}
\begin{equation}
<\hat{q}\hat{p}+\hat{p}\hat{q}>_m=\frac{-\sin 2\phi\tanh r}{1-\tanh r}
\end{equation}
\begin{equation}
<\hat{p}^2>_m=\frac{\gamma(1+\cos 2\phi\tanh r)}{2(1-\tanh r)}.
\end{equation}
The $m$ subscript denotes the expectation value with respect to the mixed
state.
These averages are not equal to equations (3.20-22) which were calculated
with respect
to the pure squeezed vacuum.
We will show in section V and VI that the differences can be important.
We can also calculate the entropy $S$. It has been shown \cite{zeh} that for
a gaussian density matrix of the form
\begin{equation}
\rho(y,z)=N\exp[-(Ay^2+iByz+Cz^2)]
\end{equation}
where $y=q-q'$ and $z=q+q'$
\begin{equation}
S=-Tr[\hat{\rho}ln\hat{\rho}]=-u^{-1}(u\ln u+v\ln v)
\end{equation}
where
\begin{equation}
u=\frac{2C^{1/2}}{A^{1/2}+C^{1/2}}, \;\;\; v=\frac{A^{1/2}-C^{1/2}}
{A^{1/2}+C^{1/2}}.
\end{equation}
In the coordinate representation the density matrix (4.5) has the form
\begin{eqnarray}
\rho(q,q')&=&N\exp\left[\frac{-\gamma }{2(1-\cos 2\phi\tanh r)}
\left(q^2(1+i\sin2\phi\tanh r)
\right.\right. \nonumber \\
&+&\left.\left.q'^2(1-i\sin 2\phi\tanh r)-2qq'\tanh r\right)\right].
\end{eqnarray}
Using this we find
\begin{equation}
u=\frac{2(1-\tanh r)^{1/2}}{(1-\tanh r)^{1/2}+(1+\tanh r)^{1/2}}, \;\;\;
v=\frac{(1+\tanh r)^{1/2}-(1-\tanh r)^{1/2}}{(1-\tanh r)^{1/2}+(1+\tanh
r)^{1/2}}.
\end{equation}

Using (4.10 and 4.13) we find that in the limit of large squeezing
$S\rightarrow 2r$
and in the small squeezing limit $S\rightarrow r$. We have doubled the result
since the field was decomposed into two infinite sets of modes. The high
squeezing limit is in agreement with those from Brandenberger et al \cite{bra2}
and Gasperini and Giovannini \cite{gas}. These authors adopted different
coarse graining schemes
which suggests that the entropy of a highly squeezed vacuum is robust
to the particular coarse graining implemented.

\section{De Sitter Phase}
Here we will specialise to a massless minimally coupled scalar field
in a de Sitter phase where $a=-1/H\eta$. For the Hamiltonians (2.10-11)
equation (3.30) becomes
\begin{eqnarray}
\ddot{g}&+&(k^2-2/\eta^2)g=0 \\
\ddot{g}_s&-&\frac{2}{\eta}\dot{g}_s+k^2 g_s=0.
\end{eqnarray}
These have the general solutions
\begin{eqnarray}
g(\eta)&=&c_1 e^{ik\eta}(1+i/k\eta) + c_2 e^{-ik\eta}(1-i/k\eta) \\
g_s(\eta)&=&c_1\eta e^{ik\eta}(1+i/k\eta) + c_2\eta e^{-ik\eta}(1-i/k\eta).
\end{eqnarray}

In order that $\mu(\eta')=-1$ as $\eta'\rightarrow -\infty$,
we take the solutions $c_2=0$. We also choose $\gamma=k$.
With these we find that (3.29) becomes
\begin{eqnarray}
\mu=\frac{-i-k^3\eta^3}{k\eta(1+k^2\eta^2)} \\
\mu_s=\frac{-k\eta(k\eta-i)}{k^2\eta^2+1}.
\end{eqnarray}
Equation (5.5) agrees with that derived by Ratra \cite{rat}.
Using (3.31-32) we find that
\begin{eqnarray}
\tanh ^2 r=\frac{1}{4k^4\eta^4+1} \\
\tanh ^2 r_s=\frac{1}{4k^2\eta^2+1},
\end{eqnarray}
and
\begin{eqnarray}
\cos 2\phi=\frac{1-2k^2\eta^2}{(1+4k^4\eta^4)^{1/2}},\;\;\;
\sin 2\phi=\frac{2k\eta}{(1+4k^4\eta^4)^{1/2}} \\
\cos 2\phi_s=\frac{-1}{(1+4k^2\eta^2)^{1/2}},\;\;\;
\sin 2\phi_s=\frac{-2k\eta}{(1+4k^2\eta^2)^{1/2}}.
\end{eqnarray}

Albrecht et al \cite{alb} and Grishchuk and Sidorov \cite{gri}
have calculated the squeeze parameter $r$ for this model
using the vacuum defined with and without the surface term respectively.
Their results agree with equations (5.7-8).

The limit of interest is $|k\eta|<<1$ which is long after Hubble crossing. This
is also the high squeezing limit. Using a standard inflation model, modes with
wavelengths of the current Hubble radius would have had $|k\eta|\approx
10^{-50}$ at the end of inflation \cite{gri}. Thus $|k\eta|<<1$ is a very good
approximation.
In this limit we find that from (5.7-8)
\begin{equation}
r_s\rightarrow -ln|k\eta|,\;\;r\rightarrow-2ln|k\eta|.
\end{equation}
This shows that, under these conditions, the vacuum defined by dropping the
surface term generates twice as much entropy.
We will show below that the increased entropy is due to enhanced
momentum fluctuations.

In the high squeezing limit we find that up to relevant order in $k\eta$
\begin{eqnarray}
\tanh r&\rightarrow& 1-2k^4\eta^4,\;\;\;\tanh r_s\rightarrow 1-2k^2\eta^2 \\
e^{i\phi}&\rightarrow& 1-k^2\eta^2/2 + i(k\eta+k^3\eta^3/2) \\
e^{i\phi_s}&\rightarrow& i-k\eta.
\end{eqnarray}
Using these  we find that (4.1) become
\begin{eqnarray}
&|&r,\phi\rangle\rightarrow N\int_{-\infty}^{\infty}\exp
\left(-y^2 k^4\eta^4\right)
|y(k\eta+k^3\eta^3/2-i(1-k^2\eta^2/2))\rangle dy \\
&|&r_s,\phi_s\rangle_s\rightarrow N\int_{-\infty}^{\infty}\exp
\left(-y_s^2 k^2\eta^2\right)|y_s(1+ik\eta)\rangle dy_s.
\end{eqnarray}
To understand the significance of (5.15-16) we must know the properties of the
physical variables for the coherent states with support in (5.15-16).
{}From (2.5) the quantized physical field is given by
\begin{equation}
\hat{\Phi}(x)=\sqrt{\frac{2}{L^3}}\sum_{\vec{k}}[\hat{Q}_{\vec{k}}^+
 \cos\vec{k}\cdot\vec{x} + \hat{Q}_{\vec{k}}^- \sin\vec{k}\cdot\vec{x}]
\end{equation}
and
\begin{equation}
\frac{d\hat{\Phi}(x)}{dt}=\sqrt{\frac{2}{L^3}}\sum_{\vec{k}}[\hat{P}_{\vec{k}}^+
 \cos\vec{k}\cdot\vec{x} + \hat{P}_{\vec{k}}^- \sin\vec{k}\cdot\vec{x}]
\end{equation}
where
\begin{equation}
\hat{Q}_{\vec{k}}^{\sigma}=\frac{\hat{q}_{\vec{k}}^{\sigma}}{a}
\end{equation}
and
\begin{equation}
\hat{P}_{\vec{k}}^{\sigma}=\frac{1}{a^2}\left(\hat{p}_{\vec{k}}^{\sigma}-
\frac{\dot{a}}{a}
\hat{q}_{\vec{k}}^{\sigma}\right)=\frac{\hat{p}_{s\vec{k}}^{\sigma}}{a^2}.
\end{equation}
The operator $\hat{Q}_{\vec{k}}^{\sigma}$ measures the amplitude of a
standing wave
of wavelength $2\pi/k$, while the operator $\hat{P}_{\vec{k}}^{\sigma}$
measures
the rate of oscillation of the wave. The canonical momenta $\hat{p}
_{\vec{k}}^{\sigma}$ and $\hat{p}_{s\vec{k}}^{\sigma}$ are defined in (2.8-9).

{}From (4.3) and (5.15-16) we have
\begin{equation}
y(k\eta+k^3\eta^3/2-i(1-k^2\eta^2/2))=\frac{1}{\sqrt{2k}}(k<q>+i<p>)
\end{equation}
\begin{equation}
y_s(1+ik\eta)=\frac{1}{\sqrt{2k}}(k<q>+i<p_s>).
\end{equation}
{}From (5.15-16) we know that after Hubble crossing the superposition
has support in the range $y=\pm 1/(k\eta)^2$ and $y_s=\pm 1/(k\eta)$.
This translates as an amplitude and canonical momenta range of
\begin{eqnarray}
&<q>&=\pm\sqrt{\frac{2}{k}}\left(\frac{1}{k\eta}+\frac{k\eta}{2}\right),\;\;\;
<p>=\pm-\sqrt{2k}\left(\frac{1}{k^2\eta^2}-\frac{1}{2}\right) \\
&<q>_s&=\pm\sqrt{\frac{2}{k}}\left(\frac{1}{k\eta}\right),\;\;\;<p_s>=
\pm\sqrt{2k}.
\end{eqnarray}
Using these and (5.19-20) we find that the physical amplitude and momentum,
for both vacua, range between
\begin{equation}
Q=\pm H\sqrt{\frac{2}{k^3}},\;\;\;P=\pm \sqrt{2k}H^2\eta^2.
\end{equation}

This result gives us a new way of interpreting the vacua of quantum
fluctuations
in the after Hubble crossing regime. It tells us that the vacua comprise a
continuous quantum superposition over coherent states (or standing waves)
with the physical amplitude and
momenta range in (5.25). Although each coherent state describes a
spatially-inhomogeneous
perturbation the total state is still spatially-homogeneous. We also see that
as time
goes on ($\eta\rightarrow 0$) the coherent states with support in the
superposition have vanishing physical momentum. This is the quantum analogue
of the freezing of classical perturbations after Hubble crossing.
The classical freezing occurs since the oscillatory factor, $e^{ik\eta}$, in
the solution to the classical equation of motion stops oscillating after
Hubble crossing since $|k\eta|<1$.
This phase
space picture is consistent with fluctuations in $Q$ and $P$ calculated using
the pure squeezed vacua which give
in the after Hubble crossing regime
\begin{equation}
(\Delta Q)^2\rightarrow \frac{H^2}{2k^3},\;\;\;(\Delta P)^2\rightarrow
\frac{H^4k\eta^4}{2}.
\end{equation}
This is true for both vacua as it must for any two pure states that differ
only by a coordinate dependent phase.

The decoherence mechanism we have proposed breaks the quantum interference
between coherent states. It is therefore of interest to see how the
fluctuations in the individual coherent
states compare with the distribution (5.25). As is well known the fluctuations
in $q$ and  canonical momenta $p,p_s$ are at the vacuum level for coherent
states. Using this and (5.19-20) we find that the fluctuations of the coherent
states are
\begin{eqnarray}
(\Delta Q)^2&=&(\Delta Q)_s^2=\frac{1}{2ka^2}=\frac{H^2\eta^2}{2k} \\
(\Delta P)_s^2&=&\frac{k}{2a^4}=\frac{1}{2}kH^4\eta^4 \\
(\Delta P)^2&=&\frac{1}{a^4}\left[(\Delta p)^2+
\frac{\dot{a}^2}{a^2}(\Delta q)^2\right]=
\frac{kH^4\eta^4}{2}\left[1+\frac{1}{k^2\eta^2}\right].
\end{eqnarray}

The first thing we notice is that the superposition bandwidth of $Q$ in (5.25)
is $1/(k\eta)$ times the vacuum fluctuation level of $Q$ calculated in (5.27).
This shows that
the $Q$ fluctuations are enhanced after Hubble crossing. It also means that
our coarse graining scheme breaks the phase space into $1/k\eta$ incoherent
pieces along the $Q$ axis. Equations (5.25) and (5.28) show that the
superposition bandwidth
in $P$ space is of the same order as the $P$ fluctuations for the
coherent states defined with
the surface term. That is, the $P$ fluctuations are unsqueezed and remain at
the vacuum level. The big difference between the two vacua can be seen by
equations (5.28) and (5.29). The coherent states defined without the surface
have fluctuations
in $P$ space of order $1/(k\eta)$ over those defined with the surface term.
Thus although
the two vacua are comprised from superpositions over coherent states with the
same range of mean values, the $P$ fluctuations of the coherent states defined
without
the surface term are of order $1/(k\eta)$ larger. This means the quantum phase
space for this vacuum is much more spread out
in the $P$ direction. Despite this
we still have the same $P$ fluctuations for both pure state vacua. However it
does suggest that if the quantum coherence is broken these enhanced $P$
fluctuations will become evident. This is indeed the case since for the
mixed state (4.5) we find
\begin{equation}
(\Delta Q)^2=(\Delta Q)_s^2\rightarrow\frac{H^2}{2k^3}
\end{equation}
\begin{equation}
(\Delta P)^2_s\rightarrow H^4k\eta^4
\end{equation}
\begin{equation}
(\Delta P)^2\rightarrow\frac{H^4\eta^2}{2k}.
\end{equation}
We see that the $Q$ fluctuations are unchanged from those derived from the
pure squeezed vacua. The $P$ fluctuations for the vacuum defined with the
surface term only differ by a factor of a half from the pure state.
However we see that the vacua defined without the surface term has
fluctuations in $P$ of order $1/(k\eta)$ larger than the other. This
shows how decoherence breaks the equivalence between the vacua. It explains the
origin of the extra entropy.

\section{astrophysical implications}
In this section we will see how the previous results have important
implications for the power spectrum of primordial fluctuations on
super-horizon scales. We will use the gauge invariant theory of
cosmological perturbations presented in \cite{bra3}. The formalism will
only be very briefly sketched here.

The action for gauge invariant cosmological perturbations
has the form
\begin{equation}
S=\frac{1}{2}\int dx^4\left[(\dot{v})^2-c_s^2\sum_{i}(v_{,i})^2+
\frac{\ddot{z}}{z}v^2-
\frac{d}{d\eta}\left(\frac{\dot{z}}{z}v^2\right)\right]
\end{equation}
where $v$ is a gauge invariant combination of metric and matter perturbations,
$c_s$ is a constant ($c_s$=1 in inflation) and $z$ is given by
\begin{equation}
z=\frac{a({\cal H}^2-\dot{{\cal H}})^{1/2}}{{\cal H}c_s}
\end{equation}
where ${\cal H}=\dot{a}/a$ is the conformal Hubble parameter (as before the
dot denotes a derivative with respect to conformal time).
The important physical quantity is the Bardeen variable $\Phi^b$. It takes the
form
\begin{equation}
\Phi^b=-\sqrt{\frac{3}{2}}l_p\frac{({\cal H}^2-\dot{{\cal H}})}{{\cal H}c_s^2}
\frac{d}{d\eta}\left(\frac{v}{z}\right)
\end{equation}
where $l_p$ is the Planck length.
The surface term
in (6.1) has been added in by hand. Albrecht et al \cite{alb} included the
same surface term for convenience. We will see that, when combined with
decoherence, the surface in (6.1) is not only convenient but necessary in
order to get a scale free spectrum.

The Lagrangian density derived from (6.1) is equivalent to (2.4) if we make the
identification
$\chi\equiv v$, $a\equiv z$ and put $\xi=m=0$. Therefore the quantization
and decoherence scheme implemented up to section V are applicable to gauge
invariant cosmological perturbations. For the de Sitter phase discussed
in section V, $z=0$ and no fluctuations
are amplified. However for a realistic inflationary scenario we would have
small deviations from the purely exponential expansion. In this case, following
the approach of Albrecht et al \cite{alb}, we can approximately put
$z(\eta)\propto a(\eta)\propto 1/\eta$.
This should be a reasonable approximation as long as we are not
interested in the overall amplitude of fluctuations. The action (6.1) is now
identical to the model discussed in section V.

Of particular interest is the power spectrum $|\delta_k|^2$, which is
the spectrum of fluctuations of the Bardeen variable $\Phi^b$.
When quantized the Bardeen variable is equivalent to (5.18) up to an overall
$k$ independent numerical factor. Therefore
we find from (5.31-32) that the power spectra has the spectral dependence
\begin{equation}
|\delta_k|_s^2\propto k,\;\;\;|\delta_k|^2\propto k^{-1}.
\end{equation}
Thus we see that, when combined with decoherence, scale invariant power
spectra (on super-horizon scales during inflation) are
only obtained if the surface term of (6.1) is included in the action.

\section{Discussion and conclusion}
So what has been learnt? First of all the coherent state representation (CSR)
has given us a new phase space representation for the quantum state of
fluctuations. As discussed in the introduction, the CSR
has advantages over the Wigner function as a phase space representation.
Work motivated by quantum optics showed that a squeezed vacuum consists
of a continous superposition over coherent states. The coherent states
with support in the superposition form a 1 dimensional line in phase space.
We showed that:
\newline
a) In the after
Hubble crossing regime this line of support rotates towards the amplitude axis
and is exponentially suppressed beyond the amplitude level in (5.25).
This shows transparently the quantum coherence
between classical-like spatially-inhomogeneous coherent states.
This quantum coherence inturn gives rise to the spatially-homogeneous squeezed
vacuum. Unlike the Wigner function the CSR shows clearly the need for
decoherence in order to generate inhomogeneities.
The coherent states with support
in the superposition have momentum that tends to zero. This is the quantum
analogue of the classical freezing of fluctuations after Hubble crossing.

By decohering the squeezed vacuum in the CSR we have a
transparent way of implementing the quantum to classical transition
that is well motivated by work on environmentally induced decoherence. We
showed
that:
\newline
b) This procedure gave the same entropy as other coarse graining methods in
the high squeezing limit. This suggests the result $S\rightarrow 2r$ for the
entropy in the high squeezing limit is robust to the coarse graining
implemented.
\newline
c) Decoherence breaks the physical equivalence between vacua that differ by a
coordinate dependent phase generated by a surface term
in the Lagrangian. This was because decoherence of the vacuum defined without
the surface term caused, in the after Hubble crossing regime, an increase in
the physical momentum
fluctuations of the order $1/(k\eta)$ over that of the corresponding pure
state. Since these enhanced
fluctuations are the origin of the extra entropy we would expect that, like
entropy, the enhancement of momentum fluctuations is a property robust to the
type of coarse graining implemented.
\newline
d) In the gauge invariant theory of cosmological perturbations, decoherence
implies that a surface term must be added to the action in order to obtain
scale invariant power spectra on super-horizon scales during an inflationary
phase.

These results suggest that great care
must be taken in choosing the surface term of time dependent systems evolving
in a non-unitary way.
For the model of section V we would conclude that keeping the surface term
is necessary if we wish to obtain physical momentum
fluctuations of the same order as that predicted by the pure state theory.
There are also technical reasons. It ensures that the double
derivative of
the scale factor does not appear in the Lagrangian. This is necessary for a
consistent variational theory when the scale factor is treated dynamically.
The physical momentum variable is of astrophysical significance.
It plays a role in the Sachs-Wolfe effect recently discussed
within the squeezed state formalism by Grishchuk \cite{gri2}.
In the context of gauge invariant cosmological perturbations
it seems that a surface term must be added to the action in order to
get results consistent with COBE \cite{smoot}. Of course the results presented
here only apply to an inflationary phase. However it seems unlikely that an
analysis that included the radiation and matter dominated era's would change
this conclusion. It may be possible to argue on theoretical grounds that
the relevant surface term should be included in the action so that no
double derivative of $z$ appears in the action.

There has been an assumption in this paper, also  implicit in
\cite{bra2}\cite{gas},
that the effect of the continuous process of decoherence can be
modelled at a given time by taking
the pure state and putting the off-diagonal terms in some chosen basis to zero.
The assumption is plausible but is by no means proved. Such a proof is
beyond the scope of this paper. A proper understanding
of the quantum to classical transition requires the introduction of an open
system.
The qualitative effect of considering an
open system is to renormalise the free system and to contribute an effective
dissipation and noise into the dynamics. The noise and dissipation are
related at a fundamental level via a fluctuation-dissipation relation.
Noise is responsible for decoherence and entropy generation. Dissipation
may have important implications for the amplitude and spectrum of
fluctuations. The effect of dissipation can not be taken into account using
the ad hoc decoherence mechanism used here and elsewhere.
Processes such as decoherence, entropy generation and dissipation in the
early universe should be studied within the rigorous framework of a
quantum field theory of open systems \cite{hu2}. However this leads to
complex non-Markovian dynamics.
A more tractable first step in this direction would be to study the dynamics
of quantum fluctuations within the framework of the quantum optical
master equation \cite{lou}. This generates Markovian dynamics. Techniques
for solving the quantum optical master equation for general time dependent
quadratic systems have recently been presented \cite{twam}.

\end{document}